\documentclass[12pt]{article}
\usepackage{bm,amsmath,amssymb,graphicx}

\begin{document}
\pagenumbering{arabic}
\begin{titlepage}

\title{Reply to ``Comment on `Quantum massive conformal gravity' 
by F. F. Faria"}

\author{F. F. Faria$\,^{*}$ \\
Centro de Ci\^encias da Natureza, \\
Universidade Estadual do Piau\'i, \\ 
64002-150 Teresina, PI, Brazil}

\date{}
\maketitle

\begin{abstract}
Recently in (Eur. Phys. J. C 76:341, 2016), Myung has suggested 
that the renormalizability of massive conformal gravity is meaningless 
unless the massive ghost states of the theory are stable. Here we show 
that massive conformal gravity can be renormalizable having unstable ghost 
states. 
\end{abstract}

\thispagestyle{empty}
\vfill
\noindent 
\bigskip
\noindent * fff@uespi.br \par
\end{titlepage}
\newpage

%%%%%%%%%%%%%%%%%%%%%%%%%%%%%%%%%%%%%%%%%%%%%%%%%%%%%%%%%%%%%%%%%%%%%%%%%%%%%%%

Before we address the content of Myung's paper \cite{Myung}, let us take a 
brief look on the results of Ref. \cite{Faria1}. For this purpose, we 
consider the massive conformal gravity (MCG) action, which is given 
by\footnote{Here we use ``absolute units" in which $c=\hbar=16\pi G=1$.}
\cite{Faria2}
\begin{equation}
S_{\textrm{MCG}} = \int{d^{4}x} \, \sqrt{-g}\bigg[ \alpha 
\big(\varphi^{2}R + 6\partial_{\mu}\varphi\partial^{\mu}\varphi\big)
 - \frac{1}{m^2}C^{\alpha\beta\mu\nu}C_{\alpha\beta\mu\nu}
\bigg],
\label{1}
\end{equation}
where $m$ is a constant with dimensions of mass, $\alpha$ is a dimensionless 
constant, $C^{\alpha}\,\!\!_{\beta\mu\nu}$ is the Weyl tensor, $R$ is the 
scalar curvature, and $\varphi$ is a scalar field called dilaton. Using the 
Lanczos identity, we can write (\ref{1})  as
\begin{equation}
S_{\textrm{MCG}} = \int{d^{4}x} \, \sqrt{-g}\bigg[ \alpha 
\big(\varphi^{2}R + 6\partial_{\mu}\varphi\partial^{\mu}\varphi\big)
- \frac{2}{m^2} \left(R^{\mu\nu}R_{\mu\nu} - \frac{1}{3}R^{2} \right) 
\bigg],
\label{2}
\end{equation}
where $R_{\mu\nu}$ is the Ricci tensor.

By performing a perturbative quantization of (\ref{2}) about the background 
field expansions
\begin{equation}
g_{\mu\nu} = \eta_{\mu\nu} + h_{\mu\nu}, 
\label{3}
\end{equation}
\begin{equation}
\varphi = \sqrt{\frac{2}{\alpha}}(1 + \sigma),
\label{4}
\end{equation}
it can be shown, after a long but straightforward calculation, that the 
Feynman propagators for the quantum fields $\sigma$ and $\Psi_{\mu\nu} 
= h_{\mu\nu} - \eta_{\mu\nu}h/2$  are given by \cite{Faria1} 
\begin{equation}
D_{\sigma} =  i\int{\frac{d^{4}p}
{(2\pi)^{4}}} \frac{e^{-ip\cdot(x-y)}}{p^2 + m^{2}- i\chi}, 
\label{5}
\end{equation} 
\begin{eqnarray}
D^{\mu\nu,\alpha\beta}_{\Psi} &=& -\frac{i}{2}\left(\eta^{\mu\alpha}
\eta^{\nu\beta} +\eta^{\mu\beta}\eta^{\nu\alpha} - \eta^{\mu\nu}
\eta^{\alpha\beta}\right) \nonumber \\ && \times \int{\frac{d^{4}p}
{(2\pi)^{4}}} \frac{m^{2}e^{-ip\cdot(x-y)}}{(p^2 - i\chi)
(p^2 + m^{2} - i\chi)},
\label{6}
\end{eqnarray}
where $\chi$ is an infinitesimal parameter. Since the residues at 
the poles for both massive terms in (\ref{5}) and (\ref{6}) are 
negative, its corresponding quantum states are taken to have negative 
energy or negative norm. In either case quantum MCG is supposed 
unphysical. However, if the states of negative residues (ghost states) are 
unstable, the theory is unitary.
 
In his paper \cite{Myung}, Myung argues that the massive ghost states 
prevents MCG from being treated perturbatively if such ghost states are 
unstable \cite{Veltman}. In this case, however, we can use a modified 
perturbation series in which only diagrams without self-energy parts for the 
unstable ghost states are included, with the bare propagators for these states 
$D(p^2)$ replaced by the dressed propagators \cite{Antoniadis}
\begin{equation}
\overline{D}(p^2) = \left[ D^{-1}(p^{2}) - \Pi(p^2) \right]^{-1},
\label{7}
\end{equation}
where $\Pi(p^2)$ is the sum of all 1PI self-energy parts. Since 
renormalization concerns only the high-energy behavior of the propagators, 
the use of dressed propagators does not affect it. So the $p^{-4}$ behavior 
of the bare propagator (\ref{6}) at high momenta makes MCG renormalizable 
regardless of whether the ghost states are unstable or not.

%%%%%%%%%%%%%%%%%%%%%%%%%%%%%%%%%%%%%%%%%%%%%%%%%%%%%%%%%%%%%%%%%%%%%%%%%%%%%%%

\end{document}